\documentclass[]{aastex701}

\usepackage{times}
\usepackage{xcolor}
\usepackage{comment}
\usepackage{amsmath}
\usepackage{graphicx}
\usepackage[caption=false]{subfig}
\usepackage{multirow}
\usepackage{rotating}
\usepackage[T1]{fontenc}
\usepackage{url}
\usepackage{pifont}
\usepackage{newtxtext,newtxmath}

\newcommand{\Ha}{H$\alpha$}
\newcommand{\Hb}{H$\beta$}

\newcommand{\OII}{[O\,\textsc{ii}]}
\newcommand{\OIII}{[O\,\textsc{iii}]}

\newcommand{\SII}{[S\,\textsc{ii}]}
\newcommand{\NII}{[N\,\textsc{ii}]}

\shortauthors{Guo et al.}

\begin{document}

\title{The Narrow-Line Seyfert 1 Phenomenon: Accretion State Versus Host Galaxy Properties}

\author{Kaizheng Guo}
\affiliation{Shanghai Astronomical Observatory, Chinese Academy of Sciences, Shanghai, 200030, China}
\affiliation{School of Astronomy and Space Science, University of Chinese Academy of Sciences, Beijing, 100049, China}
\email{kzguo@shao.ac.cn}

\author{Hassen M. Yesuf}
\affiliation{Shanghai Astronomical Observatory, Chinese Academy of Sciences, Shanghai, 200030, China}
\email{yesufh@shao.ac.cn}

\author{Lei Hao}
\affiliation{Shanghai Astronomical Observatory, Chinese Academy of Sciences, Shanghai, 200030, China}
\email{haol@shao.ac.cn}

\author{Vineet Ojha}
\affiliation{Kavli Institute for Astronomy and Astrophysics, Peking University, Beijing 100871, China}
\email{vineetojhabhu@gmail.com}

\author{Lin Lin}
\affiliation{Shanghai Astronomical Observatory, Chinese Academy of Sciences, Shanghai, 200030, China}
\email{linlin@shao.ac.cn}

\author{Zhenzhen Li}
\affiliation{Shanghai Astronomical Observatory, Chinese Academy of Sciences, Shanghai, 200030, China}
\email{lizhenzhen@shao.ac.cn}

\correspondingauthor{Hassen M. Yesuf}
\email{yesufh@shao.ac.cn}

\begin{abstract}

The physical origin of the narrow-line Seyfert 1 (NLS1) and broad-line Seyfert 1 (BLS1) dichotomy remains debated, with competing scenarios invoking host-galaxy evolution or intrinsic accretion physics. We analysed host-galaxy properties and AGN luminosities obtained from CIGALE spectral energy distribution fitting for $\sim$12,000 Type 1 AGNs from the Sloan Digital Sky Survey, of which 29\% are NLS1s. Globally, NLS1s have lower virial black hole masses, higher inferred Eddington ratios, lower stellar masses, and higher specific star formation rates than BLS1s. In the FWHM(\Hb)--$L_{\rm AGN}$ plane, the conventional 2000 km s$^{-1}$ boundary is better viewed as an empirical division within a continuous parameter space rather than a physical threshold, with Fe~II tracing the high-accretion end. 
In a host-matched subsample of 767 NLS1--BLS1 pairs with statistically indistinguishable stellar mass, black hole mass, and redshift, NLS1s still show higher Eddington ratios, stronger Fe~II emission, and bluer optical continua, together with elevated SFR and dust attenuation, suggesting that the NLS1 phenomenon is most naturally associated with a high-accretion state within the continuous distribution of Type 1 AGNs, while host-galaxy gas supply may also play a role in modulating its strength. In this picture, NLS1 and BLS1 classifications reflect different locations within a continuous accretion sequence of the same underlying population rather than two physically disjoint classes.

\end{abstract}


\section{Introduction} \label{sec:intro}

Narrow-line Seyfert 1 galaxies (NLS1s) constitute a subclass of Type 1 active galactic nuclei (AGN), originally identified by unusually narrow broad permitted emission lines (FWHM(\Hb) $< 2000$ km s$^{-1}$) and weak \OIII\ emission relative to \Hb\  \citep{osterbrock1985spectra,goodrich1989spectropolarimetry}. Within the quasar Main Sequence framework, NLS1s occupy the extreme end of Population A, characterized by strong optical Fe~II emission and high Eddington ratios \citep[$\lambda_{\rm Edd} = L_{\rm bol}/L_{\rm Edd}$;][]{boroson1992emission,sulentic2000eigenvector}. Reverberation mapping campaigns support this interpretation, showing that high-$\lambda_{\rm Edd}$ sources exhibit systematically shorter \Hb\ time lags and more compact broad-line regions than their lower accretion-rate counterparts \citep{du2018supermassive,du2019radius,dalla2020sloan}.

Yet the physical origin of the NLS1-BLS1 dichotomy remains unresolved. Two competing hypotheses dominate the literature. The first posits that NLS1s represent a distinct evolutionary stage, hosted by younger, gas-rich galaxies with elevated star formation rates \citep{mathur2000narrow,sani2010enhanced}. The second argues that NLS1s are simply high-accretion states of otherwise normal Type 1 AGNs, with spectral differences arising from intrinsic accretion physics rather than host galaxy properties \citep{boroson1992emission,sulentic2000eigenvector,ojha2020comparison}. Distinguishing these scenarios requires disentangling the respective roles of black hole mass, accretion rate, and host-galaxy properties, a challenge that has produced contradictory observational results.

The star formation properties of AGN hosts provide a key discriminant. In normal star-forming galaxies, SFR correlates tightly with stellar mass 
along the star formation main sequence \citep[SFMS;][]{brinchmann2004physical,elbaz2007reversal,renzini2015objective}. For AGN hosts, the picture is more 
complex: some studies find them preferentially above the SFMS, suggesting concurrent enhancement of star formation and AGN activity \citep{silverman2009ongoing,santini2012enhanced,zhuang2022star}, while others report AGN hosts below the SFMS, consistent with feedback-driven quenching \citep{bongiorno2012accreting,mullaney2015alma,shimizu2015decreased}.

Recent studies exemplify this tension. \citet{sani2010enhanced} reported elevated star formation indicators in NLS1s compared to BLS1s at similar AGN luminosities, supporting the evolutionary scenario. Conversely, \citet{kurian2024comparative} found no significant difference in specific star formation rates (sSFR) once stellar mass was controlled, favoring the accretion-state interpretation. These discrepancies likely stem from two fundamental limitations: (i) the difficulty of reliably separating AGN and host galaxy emission in Type 1 systems, where strong nuclear continua can bias stellar mass and star formation measurements \citep{ciesla2015constraining,cardoso2017impact}, and (ii) the failure to control for the underlying $M_{\rm BH}$-$M_{*}$ scaling relation, which drives apparent correlations between AGN and host properties \citep{kormendy2013coevolution}.

A key systematic affecting both hypotheses is that virial black hole masses derived from FWHM(\Hb) are susceptible to broad-line region (BLR) geometry and orientation effects \citep{fine2011orientation,brotherton2015orientation}. If NLS1 host galaxies are preferentially viewed at low inclinations (face-on), projected emission-line widths underestimate the true virial velocity, leading to systematically underestimated $M_{\rm BH}$ and overestimated inferred $\lambda_{\rm Edd}$ \citep{baldi2016radio,rakshit2017catalog}. A related complication arises for super-Eddington accreting massive black holes \citep[SEAMBHs;][]{wang2014supermassive,du2016supermassive}, whose radiation-pressure-modified BLR structure may further bias standard virial calibrations. These systematics do not necessarily invalidate the virial approach for statistical studies, but they motivate supplementing virial-mass-based quantities with independent, orientation-insensitive observables---such as Fe~II emission strength ($R_{4570}$) and optical continuum slope---when drawing physical conclusions about the NLS1 phenomenon.

We hypothesize that these contradictions reflect a physical decoupling: nuclear spectral properties (governed by $\lambda_{\rm Edd}$) and host galaxy properties (governed by mass) respond to different physical parameters. If correct, NLS1s and BLS1s should exhibit distinct spectral signatures even when hosted by galaxies of identical stellar mass and black hole mass. 
Conversely, if NLS1s represent a unique evolutionary stage, they should reside in systematically different host-galaxy conditions at fixed mass.
Testing this hypothesis requires robust host-AGN decomposition and controlled comparison at fixed $M_{\rm BH}$ and $M_{*}$. The low-mass regime of the $M_{\rm BH}$-$M_{*}$ plane offers a critical testbed, where NLS1s and BLS1s coexist in significant numbers. If spectral differences persist in this overlap region, they cannot be explained solely by global host mass differences, although they may still reflect a combination of accretion physics, gas supply, and line-width geometry.

In this work, we utilize host galaxy physical properties derived from CIGALE spectral energy distribution fitting by \citet{yesuf2026galaxy}, based on a parent sample of $\sim$12,000 broad-line AGN from the Sloan Digital Sky Survey \citep[SDSS;][]{liu2019comprehensive}. The CIGALE decomposition provides robust estimates of stellar mass, star formation rate, and AGN luminosity even in luminous Type 1 systems \citep{boquien2019cigale}. Our analysis proceeds in two stages: (i) statistical comparison of global host properties to establish baseline trends, and (ii) host-property-matched stacking analysis in the low-mass overlap region to isolate intrinsic nuclear differences.

This paper is organized as follows. Section \ref{sec:data} describes the sample selection. Section \ref{sec:results} presents the statistical comparison and matched-sample analysis. Section \ref{sec:discussion} examines the implications for AGN physics and co-evolution, and Section \ref{sec:conclusion} summarizes our conclusions.

\section{Data and Sample} \label{sec:data}

\subsection{Parent Sample and Spectral Properties}
Our parent sample is drawn from the comprehensive catalog of broad-line Active Galactic Nuclei (AGN) presented by \cite{liu2019comprehensive} (hereafter Liu19). This catalog was constructed from the Sloan Digital Sky Survey \citep[SDSS; ][]{york2000sloan} Data Release 7 \citep[DR7; ][]{abazajian2009seventh}, selecting objects classified as \texttt{GALAXY} or \texttt{QSO} at redshifts $z < 0.35$ to ensure coverage of the \Ha\ emission line within the SDSS spectral range (3800--9200 \AA). 

Liu19 employed an ensemble learning Independent Component Analysis \citep[EL-ICA; ][]{lu2006ensemble} technique to model and subtract the host galaxy stellar continuum, followed by multi-component fitting of the \Ha\ and \Hb\ regions to deblend narrow and broad emission lines. Broad-line AGN were identified based on the significant detection of a broad \Ha\ component, constrained by specific criteria on Full Width at Half Maximum (FWHM), signal-to-noise ratio (S/N), and flux.  This selection recovers AGN across black hole mass $M_{\rm BH} \approx 10^{5.1}$--$10^{10.3}\,M_\odot$ and Eddington ratio $\lambda_{\rm Edd} \approx 10^{-3.3}$--$10^{1.3}$, capturing the low-mass, high-accretion regime where NLS1s predominate. The final catalog contains 14,584 broad-line AGNs.

We divide the parent sample into NLS1 and BLS1 populations based solely on the width of the broad \Hb\ emission line, defined as

\begin{equation}
\begin{aligned}
\mathrm{NLS1:} \quad & \mathrm{FWHM}(\mathrm{H}\beta_{\rm broad}) < 2000\ \mathrm{km\ s^{-1}}, \\
\mathrm{BLS1:} \quad & \mathrm{FWHM}(\mathrm{H}\beta_{\rm broad}) \geq 2000\ \mathrm{km\ s^{-1}}.
\end{aligned}
\end{equation}
This yields 4,068 NLS1s and 10,516 BLS1s. We explicitly exclude the \cite{goodrich1989spectropolarimetry} \OIII/\Hb\ $< 3$ criterion, as that cut was designed for spectropolarimetric identification of hidden broad lines and introduces metallicity and ionization biases that are not central to our comparison of broad-line kinematics within a homogeneous Type 1 AGN parent sample. Accordingly, our NLS1/BLS1 nomenclature should be understood as an operational, linewidth-based classification adopted for internal consistency with Liu19, rather than as an attempt to redefine the full historical NLS1 taxonomy. In this sense, the NLS1 regime marks a part of parameter space often associated with lower virial $M_{\rm BH}$ and higher inferred $\lambda_{\rm Edd}$, rather than a fully separate AGN species \cite{sulentic2000phenomenology,boroson1992emission}.

Black hole mass is adopted directly from Liu19, who applied the virial estimator of \cite{ho2015revised}:
\begin{equation}
\log\left(\frac{M_{\rm BH}}{M_\odot}\right) = \log\left[\left(\frac{\mathrm{FWHM}(\mathrm{H}\beta)}{1000\,\mathrm{km\,s^{-1}}}\right)^2 \left(\frac{L_{5100}}{10^{44}\,\mathrm{erg\,s^{-1}}}\right)^{0.533}\right] + 6.91,
\end{equation}
where $L_{5100}$ is the continuum luminosity at 5100 \AA\ derived from spectral decomposition. 
The Eddington ratio $\lambda_{\rm Edd}$ is defined as $L_{\rm AGN}/L_{\rm Edd}$ where $L_{\rm Edd}=1.26\times10^{38}\ (M_{\rm BH}/M_\odot)\,\mathrm{erg\,s^{-1}}$. We emphasize that FWHM(\Hb), virial $M_{\rm BH}$, and $\lambda_{\rm Edd}$ are not independent quantities: $M_{\rm BH}$ is constructed from FWHM(\Hb) and luminosity, and $\lambda_{\rm Edd}$ depends explicitly on $M_{\rm BH}$. Throughout this paper, correlations involving these derived quantities are therefore interpreted as trends consistent with a high-accretion picture, rather than as fully independent causal evidence. By contrast, Fe~II strength, narrow-line ratios, continuum shape, and host-galaxy parameters provide more nearly independent constraints.

\subsection{Host Galaxy Properties}

Host galaxy physical properties are adopted from \cite{yesuf2026galaxy} (hereafter Yesuf26), who performed spectral energy distribution (SED) fitting using the Code Investigating GALaxy Emission \citep[CIGALE; ][]{boquien2019cigale}. 
Their analysis combined multi-wavelength photometry spanning ultraviolet to infrared wavelengths (GALEX, SDSS, Pan-STARRS, 2MASS, UKIDSS, WISE, IRAS 60$\micron$ when available) to decompose the observed emission into stellar, dust, and AGN components. By enforcing energy balance between absorbed ultraviolet/optical radiation and re-emitted infrared emission, CIGALE provides robust constraints on host galaxy properties even in the presence of significant AGN contamination.

We adopt the physical parameters derived from Yesuf26, using the Bayesian likelihood-weighted mean estimates for all quantities. This approach effectively marginalizes over model degeneracies by averaging across the full set of plausible SED solutions, weighted by their statistical likelihood, rather than relying on a single best-fit template. As discussed in the Supplementary Information of \citet{yesuf2026galaxy}, the CIGALE-derived values show small systematic offsets relative to previous measurements, of order $\sim$0.15 dex in $M_*$ and $\sim$0.25 dex in SFR. These offsets do not affect the main conclusions of this work, because our analysis is based primarily on relative comparisons between NLS1 and BLS1 populations within the same homogeneous measurement framework. The median uncertainties are $\sigma(\log M_*) \approx 0.1$ dex and $\sigma(\log \mathrm{SFR}) \approx 0.2$ dex. The uncertainty in AGN luminosity is luminosity-dependent and decreases toward higher $L_{\rm AGN}$; for sources with $L_{\rm AGN} > 3 \times 10^{43}\ {\rm erg\ s^{-1}}$, the characteristic uncertainty is $\sigma(\log L_{\rm AGN}) \approx 0.2$ dex.
For clarity, we distinguish below between quantities that are closer to directly measured or fitted observables---such as FWHM(\Hb), $R_{4570}$, narrow-line ratios, continuum shape, and the CIGALE-derived $M_*$, SFR, $A_V$, and $L_{\rm AGN}$---and quantities that are subsequently constructed from them, such as virial $M_{\rm BH}$, $\lambda_{\rm Edd}$, and sSFR. The latter remain useful summary parameters, but their interpretation is necessarily more model-dependent.

We cross-matched the Liu19 AGN catalog with the Yesuf26 CIGALE catalog via SDSS spectroscopic identifiers (Plate-MJD-FiberID), ensuring reliable correspondence between AGN spectral properties and host galaxy parameters. Only sources with valid measurements of both black hole mass and stellar mass were retained for further analysis.

We apply quality cuts based on CIGALE fit diagnostics to ensure the reliability of the SED-derived properties. First, we exclude sources with poor fit quality, defined as reduced chi-squared $\chi^2_\nu > 5$. Second, we require consistency between best-fit and Bayesian estimates, retaining only sources satisfying $\log( M_{*,\mathrm{best}}/M_{*,\mathrm{Bayes}}) \leq 0.5$ dex and similarly for SFR. These cuts preferentially exclude objects with complex star formation histories or uncertain AGN/host decomposition.

After applying these quality controls, our final working sample consists of 12,835 objects (3,694 NLS1s and 9,141 BLS1s), representing $\approx 90\%$ of the initial parent sample. The median $\chi^2_\nu = 0.85$ indicates well-constrained fits. From these quantities, we compute the specific star formation rate ($\mathrm{sSFR} = \mathrm{SFR}/M_*$) and the black hole-to-stellar mass ratio ($M_{\rm BH}/M_*$) for subsequent analysis.

We assessed potential selection biases that could affect the comparative analysis between NLS1 and BLS1 populations. Figure~\ref{fig:bias} presents the distributions of both subsamples in redshift space against key observational and physical properties: stellar mass ($M_*$), $V$-band absolute magnitude ($M_r$), FWHM(\Hb), and spectral signal-to-noise ratio (S/N).
Panels (a)--(c) show that both NLS1 and BLS1 populations cover similar redshift ranges ($z < 0.35$) with broadly comparable distributions in $M_*$, $M_r$, and width of Balmer emission line (FWHM(H\b)). The FWHM(\Hb) distribution in Panel (c) reflects the adopted classification boundary at 2000~km~s$^{-1}$, with NLS1s concentrated below this threshold and BLS1s extending to broader linewidths. Panel (d) shows statistically similar S/N distributions for both subsamples (median S/N $\approx 13$), indicating that there is no obvious differential degradation of spectral quality between the two classes within the analyzed sample.
At the same time, these diagnostics do not fully characterize the SDSS broad-line AGN selection function. In particular, the detectability of broad lines may still vary jointly with luminosity, linewidth, and S/N. We therefore interpret Figure~\ref{fig:bias} as evidence that no strong differential selection effect is immediately apparent in the observed distributions.

\section{Results} \label{sec:results}

\subsection{Global Demographics}

\begin{figure}[!htbp]
    \centering
    \includegraphics[width=\linewidth]{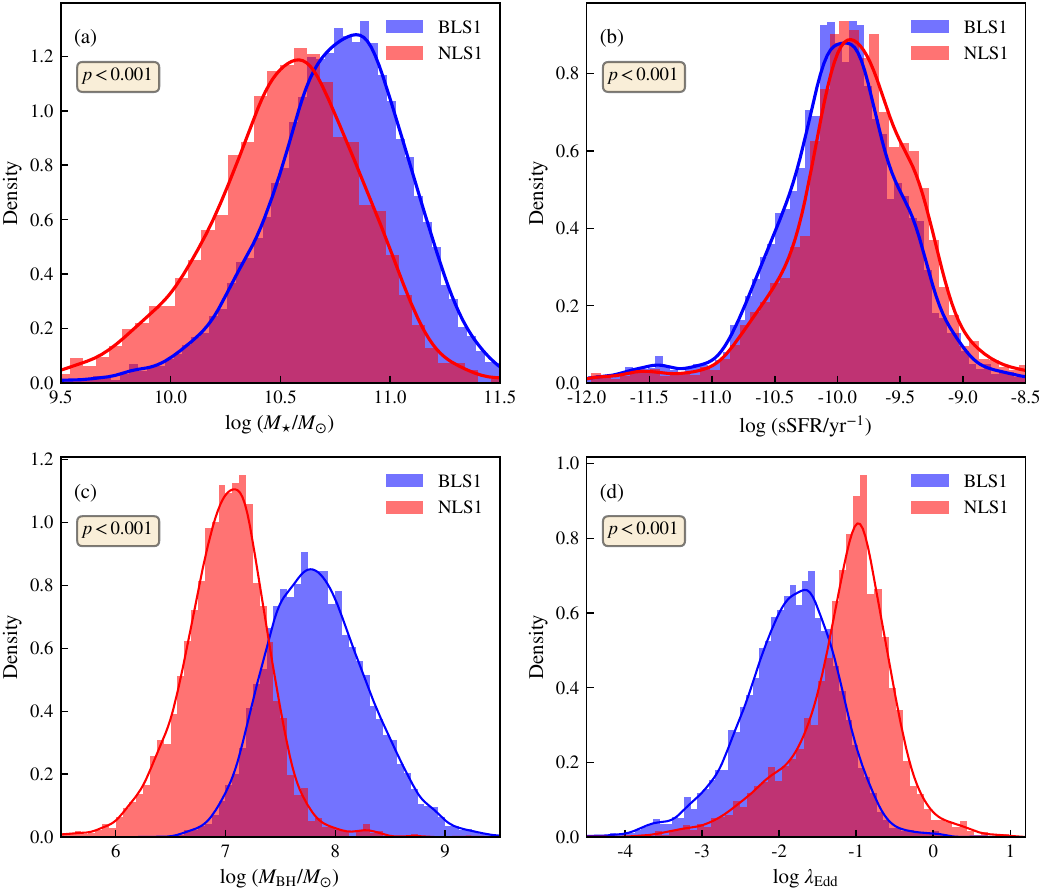}
    \caption{Distributions of (a) stellar mass, (b) specific star formation rate, (c) black hole mass, and (d) Eddington ratio for NLS1s (red) and BLS1s (blue). Gaussian kernel density estimates are overplotted as solid curves. KS test $p$-values quantify the statistical significance of differences between populations.}
    \label{fig:1}
\end{figure}

Figure~\ref{fig:1} presents the distributions of stellar mass ($M_*$), specific star formation rate (sSFR), black hole mass ($M_{\rm BH}$), and Eddington ratio ($\lambda_{\rm Edd}$) for the NLS1 and BLS1 populations. 
The stellar mass distributions (panel a) show substantial overlap but with a systematic offset: NLS1s have a mean stellar mass of $\log(M_*/M_\odot)=10.50$, with a standard deviation of 0.37 dex, whereas BLS1s have a mean of $\log(M_*/M_\odot)=10.76$, with a standard deviation of 0.33 dex. The difference is highly significant (KS test $p<10^{-16}$). This $\sim 0.2$ dex offset is consistent with the lower characteristic black hole masses of NLS1s and the $M_{\rm BH}$--$M_*$ scaling relation.
The sSFR distributions (panel b) also show a statistically significant difference in the global sample (KS test $p<10^{-16}$), with NLS1s exhibiting slightly higher sSFR on average. 

The $M_{\rm BH}$ distributions (panel c) reveal the clearest demographic separation: NLS1s have a mean black hole mass of $\log(M_{\rm BH}/M_\odot)=6.69$, with a standard deviation of 0.48 dex, while BLS1s have a mean of $\log(M_{\rm BH}/M_\odot)=7.43$, with a standard deviation of 0.60 dex (KS test $p<10^{-16}$). It is important to note that the $L_{\rm AGN}$ and $L_{5100}$ distributions differ only modestly between the two populations. Because the BLR radius scales approximately as $R_{\rm BLR} \propto L^{1/2}$, these small luminosity offsets imply only a minor difference in characteristic BLR size. Therefore, the $\sim0.7$ dex offset in virial $M_{\rm BH}$ is driven primarily by the kinematic linewidth term ($M_{\rm BH} \propto \rm FWHM^2$) rather than by a substantial difference in BLR radius. This point also motivates caution: part of the NLS1/BLS1 separation in virial mass may reflect broad-line geometry and orientation, not only intrinsic black hole mass.

The $\lambda_{\rm Edd}$ distributions (panel d), calculated from the CIGALE-derived AGN luminosities and virial black hole masses as described above, show the most pronounced difference. The NLS1 distribution is systematically shifted toward higher inferred Eddington ratios and is more concentrated in the high-$\lambda_{\rm Edd}$ regime than that of BLS1s. NLS1s have a mean of $\log \lambda_{\rm Edd}=-1.04$, with a standard deviation of 0.70 dex, whereas BLS1s have a mean of $\log \lambda_{\rm Edd}=-1.45$, with a standard deviation of 0.68 dex. A K-S test performed on these distributions indicates a highly statistically significant difference between the Eddington-ratio distributions of the NLS1 and BLS1 samples ($p<10^{-16}$). This result is consistent with numerous previous studies \citep{rakshit2017optical,ojha2020comparison,ojha2025relative}. Nevertheless, because $\lambda_{\rm Edd}$ depends explicitly on virial $M_{\rm BH}$, part of this contrast is mathematically coupled to the linewidth-based classification itself; for that reason, we rely more heavily on Fe~II strength, continuum shape, and matched-sample behavior when drawing physical conclusions.

\begin{figure}[!htbp]
    \centering
    \includegraphics{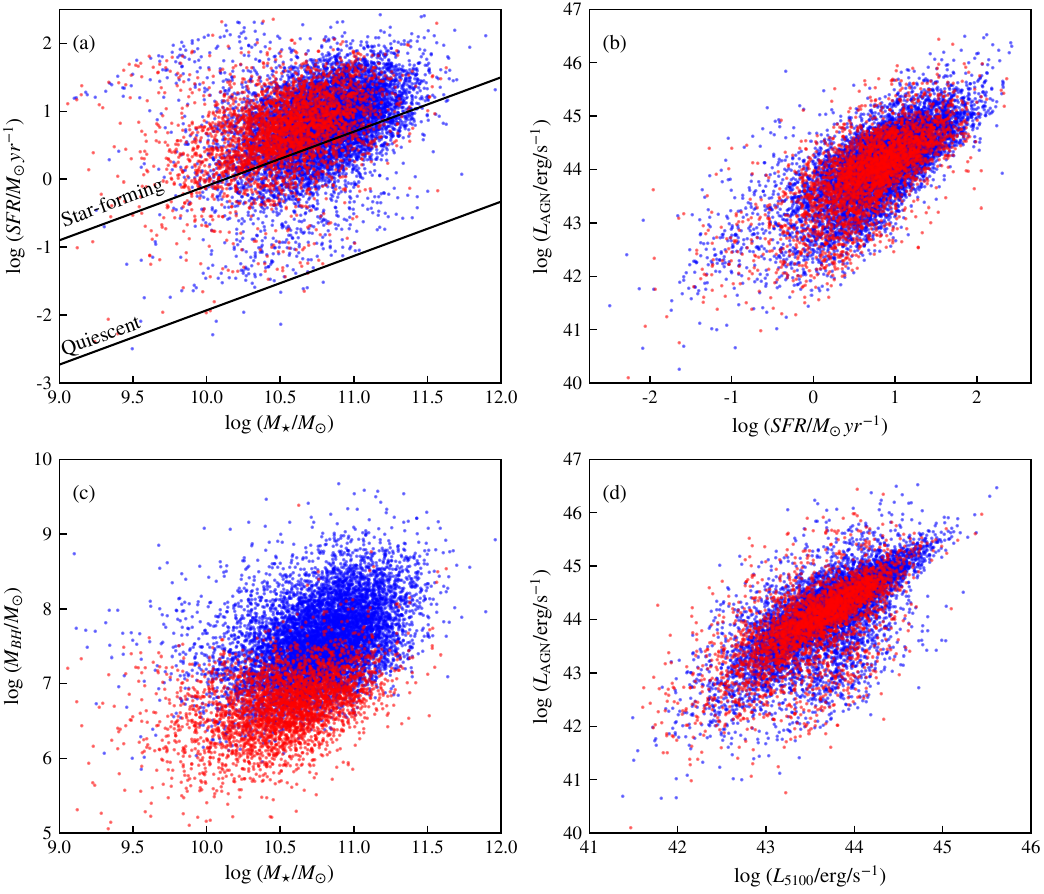}
    \caption{(a) $M_*$ versus SFR, with the SFMS of Yesuf26 shown as solid lines. (b) $L_{\rm AGN}$ versus SFR. (c) $M_{\rm BH}$ versus $M_*$. (d) SED-derived $L_{\rm AGN}$ versus continuum luminosity $L_{5100}$ from spectral decomposition. NLS1s in red, BLS1s in blue.}
    \label{fig:2}
\end{figure}

Figure~\ref{fig:2} examines the relationships between AGN properties and host galaxy characteristics. Panel (a) shows the $M_*$-SFR plane, with the \citet{yesuf2026galaxy} star-forming main sequence (SFMS) overplotted for reference. Both NLS1 and BLS1 hosts are predominantly star-forming: 75\% of NLS1s and 77\% of BLS1s lie on or above the SFMS. The distribution of our sample galaxies is broadly consistent with several previous studies which found that AGN host galaxies are often located on or above the SFMS \citep{mullaney2015alma,kakkad2017alma,koss2021bat,koutoulidis2022host}.
Panel (b) reveals a strong positive correlation between AGN luminosity and star formation rate (Spearman $\rho = 0.73$ for NLS1s, 0.71 for BLS1s). This $L_{\rm AGN}$-SFR correlation is consistent with a common gas supply fueling both star formation and black hole accretion, although both quantities correlate with stellar mass and redshift.

The relationship between black hole mass ($M_{\rm BH}$) and host galaxy stellar mass ($M_*$) is one of the key scaling relations in galaxies and is widely regarded as observational evidence for the co-evolution of galaxies and their central supermassive black holes (SMBHs). Panel (c) shows the $M_{\rm BH}$-$M_*$ relation. Both populations follow the expected positive correlation, but NLS1s systematically occupy the lower envelope at fixed $M_*$, especially apparent at higher stellar masses, where BLS1 galaxies tend to host significantly more massive black holes. This separation is expected given the virial mass estimator and the linewidth-based NLS1/BLS1 definition, and therefore should not by itself be interpreted as evidence for a distinct physical class.

Panel (d) compares $L_{\rm AGN}$ with $L_{5100}$, the continuum luminosity at 5100~\AA\ from spectral decomposition. The tight correlation ($\rho \approx 0.9$) validates the consistency between the SED-derived $L_{\rm AGN}$ and the Liu19 spectral measurements.

\subsection{Continuous Variation Across the FWHM-$L_{\rm AGN}$ Plane}

\begin{figure}[!htbp]
    \centering
    \includegraphics[width=0.8\textwidth]{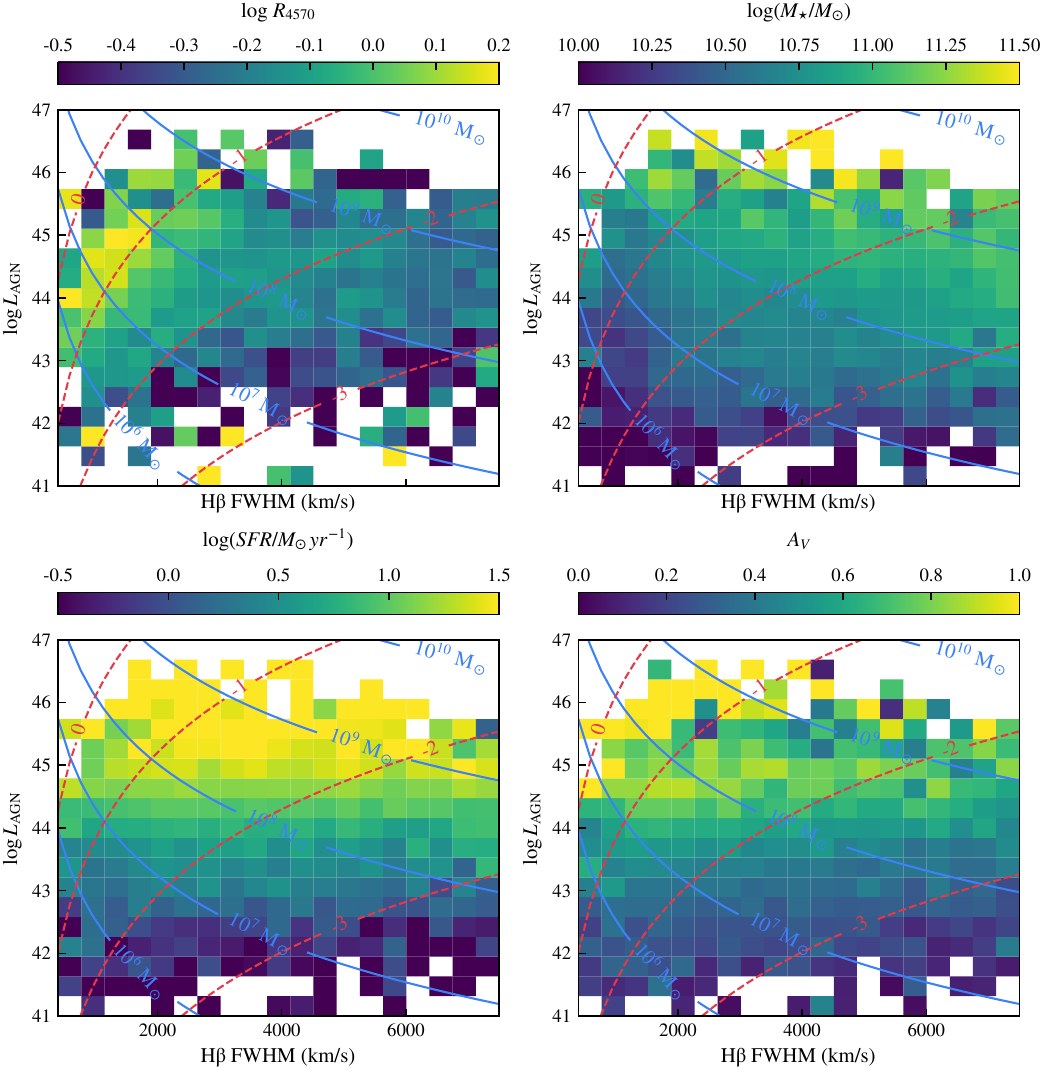}
    \caption{Two-dimensional binned statistics in the FWHM(\Hb)-$L_{\rm AGN}$ plane. Color scales show median values of (top left) log $R_{4570}$, (top right) log $M_*$, (bottom left) log SFR , and (bottom right) $A_V$. Blue solid contours indicate constant $M_{\rm BH}$; red dashed contours indicate constant $\log \lambda_{\rm Edd}$.}
    \label{fig:3}
\end{figure}

\begin{figure}[!htbp]
    \centering
    \includegraphics[width=.8\textwidth]{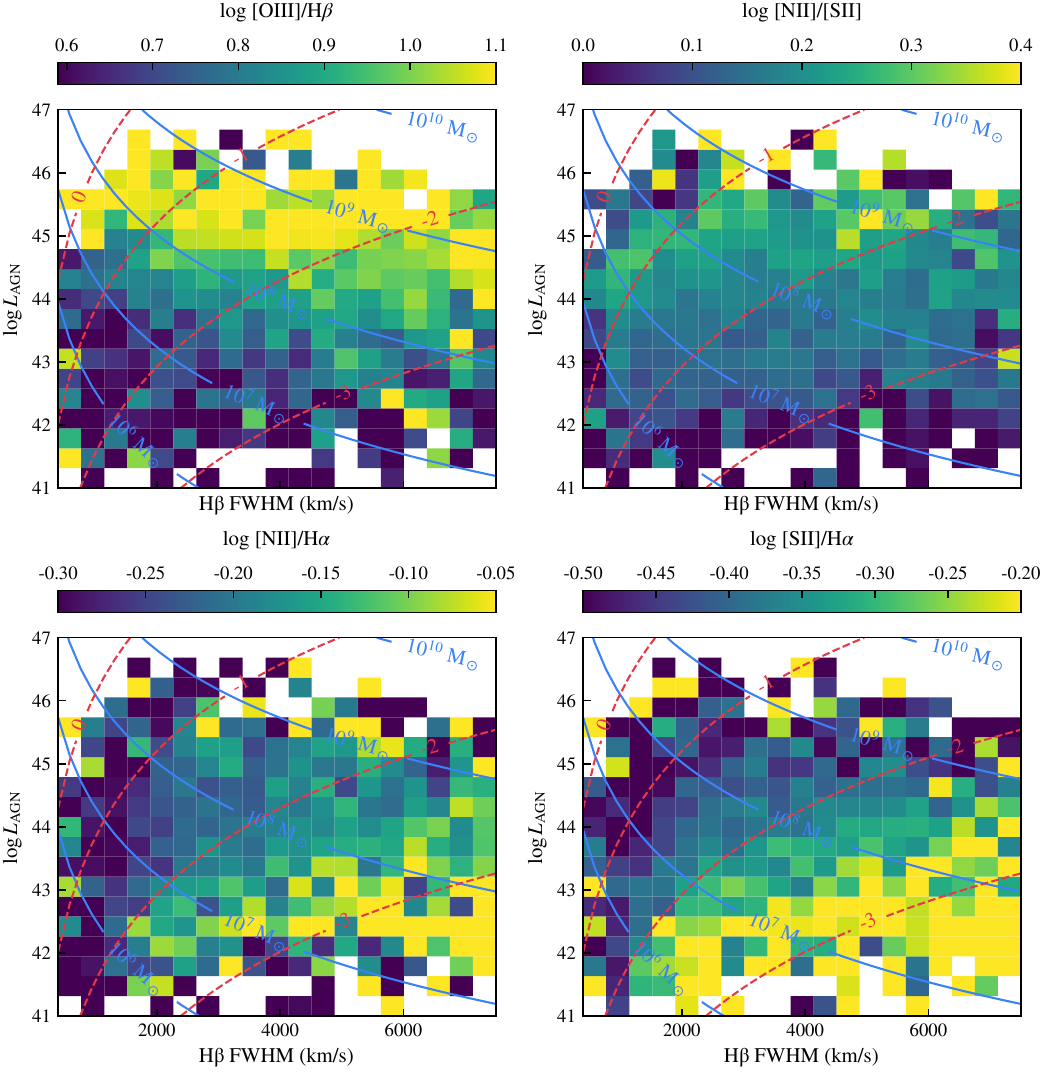}
    \caption{Same as Figure~\ref{fig:3}, but for narrow-line ratios: log(\OIII/\Hb) (top left), log(\NII/\SII) (top right), log(\NII/\Ha) (bottom left), and log(\SII/\Ha) (bottom right)}
    \label{fig:4}
\end{figure}

To assess whether the NLS1/BLS1 dichotomy represents a physically meaningful boundary or merely samples different regions of a continuous distribution, Figures~\ref{fig:3} and \ref{fig:4} present two-dimensional binned statistics in the plane of FWHM(\Hb) versus $L_{\rm AGN}$. We overlay contours of constant $M_{\rm BH}$ (blue solid) and $\lambda_{\rm Edd}$ (red dashed) computed from the virial scaling formalism.
Figure~\ref{fig:3} shows host galaxy and BLR properties. 
Fe~II strength $R_{4570}$ (top-left): The strongest gradient aligns with $\lambda_{\rm Edd}$ contours, not $M_{\rm BH}$. $R_{4570}$ increases systematically from log $R_{4570} \approx -1$ at log $\lambda_{\rm Edd} \approx -2.5$ to log $R_{4570} \approx 0$ at $\lambda_{\rm Edd} \approx 1$. This is consistent with the classical Eigenvector 1 sequence \citep{boroson1992emission}, with NLS1s occupying the high-$R_{4570}$, high-$\lambda_{\rm Edd}$ regime.
stellar mass $M_*$ (top right): The gradient follows $M_{\rm BH}$ contours, consistent with the $M_{\rm BH}$-$M_*$ relation. At fixed $M_{\rm BH}$ (following a blue contour), $M_*$ varies little across the FWHM boundary, indicating that host galaxy mass alone is unlikely to explain the NLS1/BLS1 spectral distinction.
star formation rate SFR (bottom left): SFR increases primarily with $L_{\rm AGN}$ (vertical gradient), with weak dependence on FWHM at fixed luminosity. This behavior suggests that star formation is more closely tied to global luminosity and host scaling relations than to the empirical linewidth boundary itself.
Dust attenuation $A_V$ (bottom right): Similar to SFR, $A_V$ correlates with $L_{\rm AGN}$ but shows no sharp transition at $\rm FWHM = 2000\ km\ s^{-1}$.

Figure~\ref{fig:4} presents narrow-line region diagnostics in the $L_{\rm AGN}$ versus \Hb\ FWHM plane. The ratio of \OIII/\Hb\ (top-left) varies primarily with $L_{\rm AGN}$ (vertical gradient), nearly orthogonal to the $\lambda_{\rm Edd}$ trend. The anti-correlation with $R_{4570}$ is the classic Eigenvector 1 relation.
In contrast, the ratio of \NII/\SII\ (top-right), a tracer of gas-phase metallicity, shows relatively small variations across the parameter space and no alignment with the FWHM boundary, suggesting a lack of strong metallicity gradients within the sample. However, both the \NII/\Ha\ (bottom-left) and \SII/\Ha\ (bottom-right) exhibit a distinct distribution, with higher values concentrated in the lower-right part of the diagrams, corresponding to the regime of low Eddington ratios. 
The absence of sharp transitions in host properties or narrow-line ratios at $\rm FWHM = 2000\ km\ s^{-1}$ suggests that this boundary is not a hard physical threshold. A more cautious interpretation is that it acts as an empirically useful divider that preferentially selects the high-accretion end of a continuous Type 1 AGN sequence, with FWHM serving as an imperfect and potentially orientation-dependent proxy for accretion state.

To further quantitatively validate the continuous nature of the AGN population across the $\mathrm{FWHM} = 2000\ \mathrm{km\ s^{-1}}$ boundary, we examine the direct correlations between FWHM(\Hb) and key physical drivers (see Appendix Figure~\ref{fig:fwhm_correlations}). For the combined sample, FWHM(\Hb) exhibits a strong anti-correlation with the Eddington ratio ($\rho = -0.76$, $p < 10^{-16}$). However, this tight relation is partially driven by mathematical coupling, as $\lambda_{\rm Edd} \propto M_{\rm BH}^{-1} \propto \mathrm{FWHM}^{-2}$. A more robust, independent physical test is the relationship between FWHM(\Hb) and the Fe~II strength ($R_{4570}$). For the combined sample, a significant anti-correlation is observed ($\rho = -0.39$, $p < 10^{-16}$), consistent with the classic Quasar Main Sequence framework. When restricting the statistical analysis to the NLS1 subsample alone, this correlation becomes substantially weaker ($\rho = -0.16$), indicating that the strong global trend is driven primarily by the broader continuous sequence traced jointly by NLS1s and BLS1s rather than by a strong monotonic trend internal to the NLS1 population itself. This behavior supports the view that NLS1s represent the high-accretion end of a continuous Type 1 AGN distribution rather than a physically isolated class.

\subsection{Controlled Comparison: Mass-Matched Subsample} \label{sec:matched}

The overlap region at $\log M_* < 10.8$ and $\log M_{\rm BH} < 7.5$ contains NLS1s and BLS1s with statistically indistinguishable host and black hole masses. This provides a useful matched sample in which we can ask whether the nuclear differences between the two subsamples persist after controlling for $M_*$, $M_{\rm BH}$, and $z$. 

\begin{figure}[!htbp]
    \centering
    \includegraphics{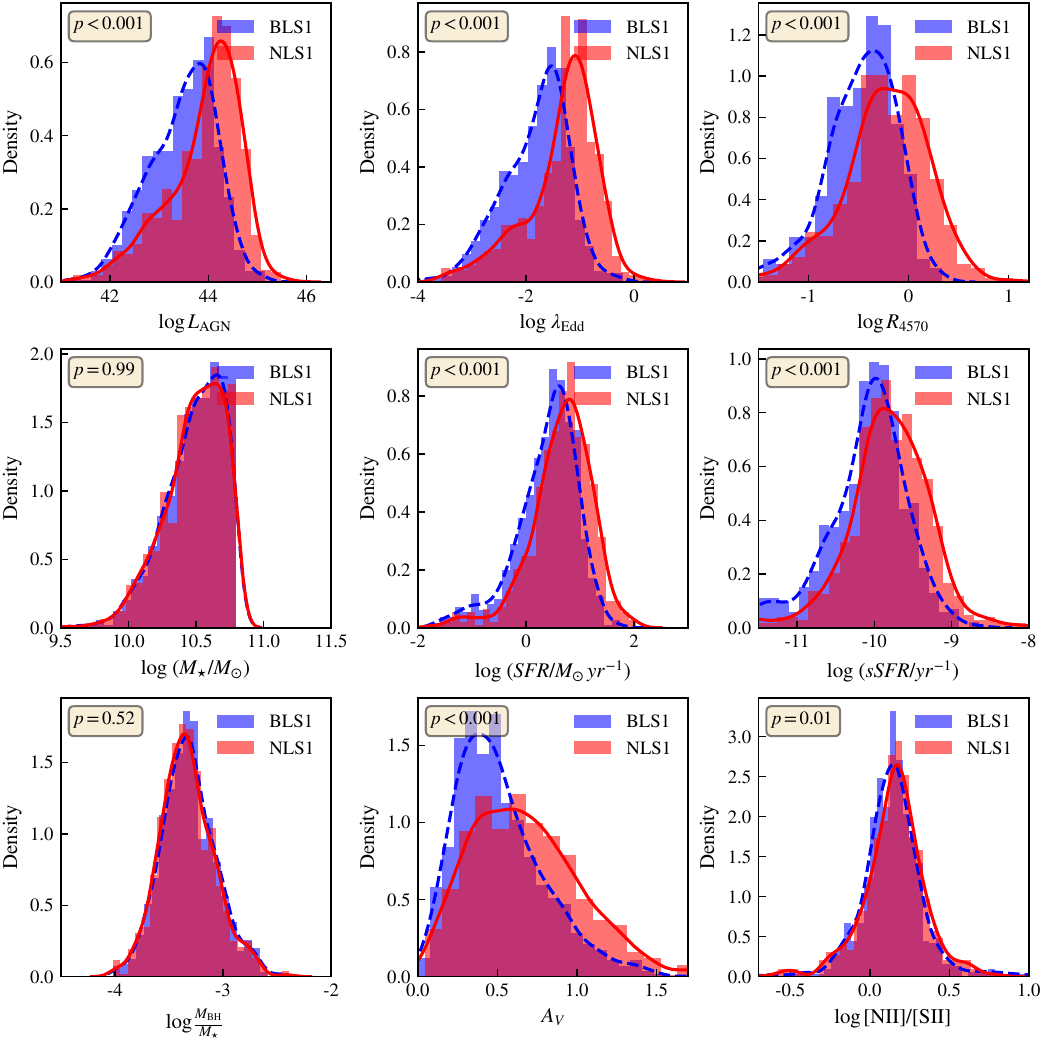}
    \caption{Comparison of matched NLS1 (red) and BLS1 (blue) subsample by $M_*$, $M_{\rm BH}$, and $z$. Top row: AGN properties ($L_{\rm AGN}$, $\lambda_{\rm Edd}$, $R_{4570}$). Middle row:Host properties ($M_*$, SFR, sSFR). Bottome row: $M_{\rm BH}/M_*$, $A_{\rm V}$ and \NII/\SII. Solid curves show kernel density estimates. KS test $p$-values are indicated in each panel.}
    \label{fig:5}
\end{figure}

We construct matched subsamples using k-d tree nearest-neighbor matching in $(\log M_*, \log M_{\rm BH}, z)$ space, yielding 767 NLS1-BLS1 pairs. Each source is matched to its nearest neighbour in the standardized three-dimensional space, subject to a maximum Euclidean distance of 0.3, corresponding approximately to $|\Delta \log M_*| \lesssim 0.09$~dex, $|\Delta \log M_{\rm BH}| \lesssim 0.10$~dex, and $|\Delta z| \lesssim 0.02$ simultaneously. The median residuals in the matched pairs for $\Delta \log M_*$, $\Delta \log M_{\rm BH}$, and $\Delta z$ are all nearly consistent with zero, confirming tight matching across all three control variables. Figure~\ref{fig:5} compares their distributions. The matching efficacy is confirmed by KS test $p$-values of 0.99 for $M_*$ and 0.52 for $\log(M_{\rm BH}/M_*)$, indicating that the matched NLS1 and BLS1 subsamples are statistically indistinguishable in the primary control variables. The purpose of this procedure is to reduce known global confounding from $M_*$, virial $M_{\rm BH}$, and $z$. As a robustness check, we also repeated the matching including SFR and $A_V$ as additional control variables, and found that the main differences in AGN properties remain, although this is confirmed with a smaller matched subsample.

With these confounding variables controlled, significant differences persist. At fixed host and black hole mass, NLS1s show higher $L_{\rm AGN}$ (median $\Delta \log L_{\rm AGN} = 0.55$ dex, $p<0.001$) and, consequently, higher inferred $\lambda_{\rm Edd}$ (median $\Delta \log \lambda_{\rm Edd} = 0.50$ dex, $p<0.001$). Their Fe~II strength also remains significantly enhanced ($\Delta \log R_{4570} = 0.32$ dex, $p<0.001$). Among these quantities, $R_{4570}$ and the continuum differences discussed below provide nearly independent support for a high-accretion interpretation, whereas the $\lambda_{\rm Edd}$ contrast must be read in the context of its definition coupling to FWHM and virial $M_{\rm BH}$.
In contrast, host-galaxy differences are more modest but not negligible: NLS1 hosts exhibit slightly higher SFR (median $\Delta \log \rm SFR = 0.23$ dex, $p<0.001$), slightly higher sSFR ($\Delta \log \rm sSFR = 0.24$ dex, $p<0.001$), and somewhat larger dust attenuation ($\Delta A_{\rm V} = 0.17$ mag, $p<0.001$). The \NII/\SII\ ratio shows only a marginal difference ($p=0.01$), with NLS1s slightly elevated, possibly reflecting differences in ionization conditions or gas-phase abundance. These residual host differences argue against the strongest possible claim that host-galaxy conditions are irrelevant; instead, they suggest that host-scale gas supply and dusty environment likely play a secondary role.

\begin{figure}[!htbp]
    \centering
    \includegraphics{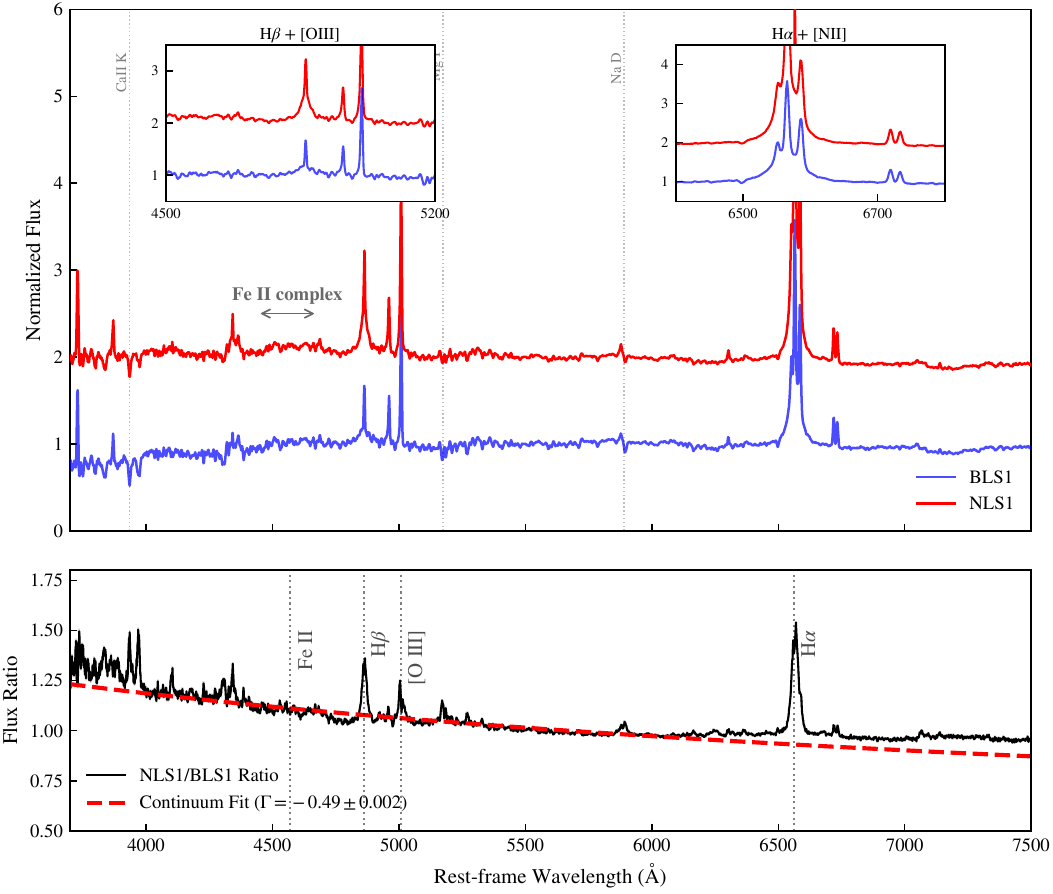}
    \caption{Top: Inverse-variance-weighted composite rest-frame spectra of the matched NLS1 (red) and BLS1 (blue) subsamples ($N=767$ in each class). The spectra are normalized over 5500--5600~\AA, and a constant vertical offset is applied to the NLS1 composite for visual clarity. The left inset shows the \Hb\ + \OIII\ region (4500--5200~\AA), and the right inset shows the \Ha\ + \NII\ region (6400--6800~\AA). Vertical dotted lines mark the principal spectral features. Bottom: Flux ratio between the NLS1 and BLS1 composite spectra (black), together with the best-fitting power-law continuum (red dashed; $\Gamma = -0.49$). The ratio spectrum highlights the bluer optical continuum of NLS1s and their enhanced Fe~II emission over 4000--5500~\AA.}
    \label{fig:6}
\end{figure}

To verify that the statistical differences identified above have direct spectral counterparts, we construct inverse-variance-weighted composite rest-frame spectra for the 767 matched NLS1s and BLS1s. Figure~\ref{fig:6} shows the composite spectra together with the NLS1/BLS1 flux ratio, while the broad--narrow decomposition of the stacked \Hb\ profiles is shown in Appendix Figure~\ref{fig:hb_fit}. The ratio spectrum is well described by a power-law continuum, indicating that NLS1s are systematically bluer than BLS1s at fixed $M_*$ and $M_{\rm BH}$. In addition, the broad excess at 4000--5500~\AA\ demonstrates stronger optical Fe~II emission in the NLS1 composite, while the weaker \OIII\ emission further reinforces the classic Eigenvector 1 trend. These composite spectra therefore confirm that the population-level differences seen in Figure~\ref{fig:5} reflect genuine spectral distinctions rather than statistical scatter in the matched samples. They also show that the key nuclear differences survive the matching. %even though some host differences remain.

\section{Discussion} \label{sec:discussion}

Our global sample analysis reveals that both NLS1 and BLS1 hosts predominantly reside on or above the SFMS, with 75\% of NLS1s and 77\% of BLS1s lying in this region. This high occupancy of NLS1s and BLS1s on or above the SFMS is consistent with a picture in which AGN activity is fueled by the same cold-gas reservoirs that sustain star formation in normal star-forming galaxies \citep{yesuf2020gas, ward2022cosmological}. We also find a statistically significant positive correlation between AGN luminosity and SFR for both populations ($\rho \approx 0.7$). This trend is consistent with a common gas-supply interpretation, but by itself it does not establish a direct physical coupling, because both quantities are expected to covary with stellar mass, redshift, and gas content. To separate the contribution of shared stellar-mass dependence from a genuine AGN--SFR link, we compute the partial Spearman rank correlation coefficient between $\log L_{\rm AGN}$ and $\log \mathrm{SFR}$ after controlling for $\log M_*$. The partial correlation coefficient is $\rho_{\rm partial} = 0.67$ ($p < 10^{-16}$) for the full sample. When further controlling jointly for both $\log M_*$ and $z$, the partial coefficient remains $\rho_{\rm partial} = 0.47$ ($p < 10^{-16}$). The persistence of a highly significant partial correlation after removing stellar-mass and redshift trends shows that the $L_{\rm AGN}$--SFR connection is not explained by stellar mass alone. More cautiously, it suggests that AGN activity and star formation are still linked in a way that is consistent with a shared cold-gas supply.%, although the present analysis does not by itself identify the underlying physical mechanism. 

Our finding that the SFR-$L_{\rm AGN}$ relation is broadly similar for NLS1s and BLS1s differs from the trend reported by \citet{sani2010enhanced}, who found stronger star-formation signatures in NLS1s at fixed AGN luminosity. One likely reason is that \citet{sani2010enhanced} used the 6.2~$\mu$m PAH feature relative to the AGN mid-infrared continuum, which primarily traces circumnuclear star formation within the central few kiloparsecs, whereas our CIGALE-based measurements are derived from multi-wavelength SED decomposition and more closely reflect the global host-galaxy SFR. The discrepancy may therefore arise because the two studies probe different physical scales. In addition, PAH-based measurements can be affected by continuum dilution and the local radiation field, whereas SED decomposition is designed to separate stellar, dust, and AGN components in a more global sense.

Recent work employing CIGALE SED fitting likewise reported no significant differences in the distributions of $M_{*}$ or sSFR between NLS1 and BLS1 hosts \citep{kurian2024comparative}. Our results, however, reveal a more nuanced picture. In our global sample, BLS1s are significantly more massive, while NLS1s exhibit systematically higher sSFRs. This discrepancy with previous literature may arise from differences in sample selection, redshift coverage, or details of the adopted SED modeling. More importantly, our matched-sample results suggest that the relevant question is not whether host properties matter at all, but whether they remain the primary driver once black hole and host masses are controlled; our answer is that they do not disappear, yet neither do they erase the nuclear spectral distinction.

There exists a continuous distribution of optical line widths in Seyfert 1 galaxies, and the separation between NLS1s and BLS1s at $\rm FWHM = 2000~km~s^{-1}$ does not correspond to an obvious physical discontinuity. Historically, the identification of objects with narrow Balmer lines that exhibit weak Fe~II and strong \OIII\ has challenged a rigid classification \citep{grupe1999new}. This is consistent with the parameter space shown in Figure~\ref{fig:3} and Figure~\ref{fig:4}, where the $\lambda_{\rm Edd}$ and $M_{\rm BH}$ contours span both the NLS1 and BLS1 regions without a clear break. At the same time, the boundary is not devoid of observational meaning: it remains an empirically useful way to isolate a population that is, on average, more Fe~II-strong and more consistent with high accretion. In this sense, FWHM acts as a practical but imperfect classifier within a continuous sequence, not as a marker of physically distinct populations.

Our comparative study of the host galaxy and spectral properties, particularly within the low-mass overlap regime, provides new constraints on the physical nature of the NLS1 phenomenon. Historically, a prevalent hypothesis has been that NLS1s represent a distinct evolutionary class of primordial AGN, potentially hosted by younger, gas-rich, and more actively star-forming galaxies compared to BLS1s \citep{mathur2000narrow, sani2010enhanced}. Our analysis of the mass-matched sample suggests a more intermediate picture. When controlling for black hole mass and stellar mass, the host properties would be expected to converge completely if host structure were irrelevant; yet they would be expected to remain the dominant separator if the NLS1 phenomenon were purely host-driven. Instead, we find that NLS1 hosts exhibit significantly higher sSFR than their BLS1 counterparts ($p<0.001$), while the strongest residual differences still lie in the nuclear spectral domain. This implies that NLS1s are more likely to occur in hosts with more active gas processing at fixed mass, but that host conditions alone do not account for the full spectral distinction.

If the host masses are similar but the star formation efficiencies differ, the origin of the distinct spectral features of NLS1s is most naturally linked to the central engine accretion state. Our spectral stacking analysis provides quantitative support for this interpretation. Even after strictly matching the NLS1 and BLS1 samples in $M_{*}$ and $M_{\rm BH}$, the ratio spectrum (Figure~\ref{fig:6}) reveals that NLS1s possess a systematically bluer optical continuum and stronger Fe~II emission. This intrinsic spectral difference aligns well with the Eigenvector 1 framework, in which accretion state is thought to be a primary driver of quasar spectral diversity \citep{boroson1992emission,sulentic2000eigenvector,marziani2018main}. In our sample, NLS1s also exhibit significantly higher inferred $\lambda_{\rm Edd}$ than BLS1s at fixed black hole mass, but the stronger physical weight should be assigned to the more direct observables---continuum shape and Fe~II strength---rather than to $\lambda_{\rm Edd}$ alone.
Standard accretion disk theory predicts that a higher accretion rate leads to a higher maximum disk temperature \citep{shakura1973black}, shifting the spectral energy distribution peak toward shorter wavelengths and producing the observed spectral hardening. This provides a plausible physical explanation for the bluer optical continua of NLS1s in the matched sample.

A separate question is whether the narrower broad \Hb\ widths in NLS1s can be attributed primarily to BLR size or, alternatively, to BLR geometry and orientation. 
The relatively small offsets in $L_{\rm AGN}$ and $L_{5100}$ between NLS1s and BLS1s imply that luminosity alone would not predict a large systematic difference in BLR size under the standard radius--luminosity relation\citep{bentz2013low}.
Reverberation-mapping studies of high-accretion AGNs suggest that BLR size depends not only on luminosity but also on accretion rate, with some SEAMBHs showing shortened \Hb\ lags at fixed $L_{5100}$, although others remain broadly consistent with the canonical relation \citep{hu2015supermassive,hu2021supermassive,huang2019reverberation}. 
The narrower broad \Hb\ widths in NLS1s are therefore unlikely to be explained simply by larger BLR sizes. They are more plausibly linked to a combination of accretion-dependent changes in BLR structure and geometric projection effects. In the quasar main-sequence framework, NLS1s occupy the extreme Population A regime, where strong Fe~II emission and relatively narrow \Hb\ widths are associated primarily with high Eddington ratio, while orientation contributes additional scatter---and potentially part of the systematic offset---to the observed line width \citep{decarli2008black,shen2014diversity,marziani2018main}. If the BLR is flattened, sources viewed more face-on can appear artificially narrow, causing virial $M_{\rm BH}$ to be underestimated and inferred $\lambda_{\rm Edd}$ to be overestimated.
Several studies have found statistical evidence for orientation-dependent line width effects in Seyfert~1 samples and NLS1s specifically \citep{fine2011orientation,brotherton2015orientation,baldi2016radio,liu2016properties,rakshit2017catalog}. However, this scenario is inconsistent with the observed differences in continuum slope ($\Gamma = -0.49 \pm 0.002$) and Fe~II emission ($\Delta \log R_{4570} = 0.32$~dex) in the matched composites: both are orientation-insensitive photometric and spectroscopic properties that are not constructed from line-width geometry, yet they remain significantly enhanced in matched NLS1s. Furthermore, the residual elevation in sSFR in matched NLS1 hosts---present even after controlling for $M_*$ and $M_{\rm BH}$---is most naturally explained by a genuinely higher gas supply fueling both accretion and star formation, rather than by a viewing-angle effect. The most defensible statement is that orientation likely contributes scatter to the linewidth distribution and introduces uncertainty of order $\sim$0.2~dex into individual virial masses \citep{jarvis2006orientation}, but it does not, by itself, account for the persistence of bluer continua, stronger Fe~II emission, and residual host differences in the matched sample.

Our results fit naturally within this picture: the alignment of $R_{4570}$ with the high-accretion end of the sequence, together with the continuity of host-galaxy properties across the conventional FWHM boundary, suggests that the NLS1 phenomenon reflects one extreme of a continuous Type 1 AGN sequence rather than a physically distinct class.
Appendix Figure~\ref{fig:corrs_comp} compares the correlations of $R_{4570}$, $M_*$, and sSFR with $\lambda_{\rm Edd}$, $L_{\rm AGN}$, and $M_{\rm BH}$. Among these quantities, $R_{4570}$ shows the strongest dependence on $\lambda_{\rm Edd}$ relative to either $L_{\rm AGN}$ or $M_{\rm BH}$, consistent with \citet{ojha2026unraveling}. This trend persists within the NLS1 subsample alone.
This interpretation is also consistent with the direct FWHM(\Hb)-$\lambda_{\rm Edd}$ and FWHM(\Hb)-$R_{4570}$ correlations shown in Appendix Figure~\ref{fig:fwhm_correlations}, which place the NLS1 population on the same underlying Eigenvector 1 sequence as the broader Type 1 AGN population.

We propose a more cautious evolutionary scenario in which NLS1s represent a preferred high-accretion spectral phase within the duty cycle of low-to-moderate mass supermassive black holes. In the $M_{\rm BH}$-$M_{*}$ plane, NLS1s are located systematically below the sequence defined by massive BLS1s, which resembles the behavior of growing black holes that are catching up to the local scaling relations \citep{kormendy2013coevolution, reines2015relations}. 
Historically, NLS1s have been characterized as the low-mass, high-accretion counterparts of classical BLS1s, typically hosting black holes with masses $M_{\rm BH} \sim 10^{6}$--$10^{7}\,M_\odot$ \citep{grupe2004mbh,williams2018narrow}. Our findings refine this picture: while NLS1s do occupy the low-mass regime globally, our mass-matched analysis reveals that they can coexist with BLS1s at identical $M_{\rm BH}$ and $M_{*}$ in the overlap region. The distinction between classes at fixed mass is therefore governed primarily by nuclear spectral state, while host gas supply and orientation likely modulate where a given source appears within that sequence.

It is plausible that secular processes or minor mergers supply the gas necessary to trigger near-Eddington accretion episodes, manifesting as the NLS1 spectral type \citep{komossa2007narrow,jarvela2018near}. The concurrent sSFR enhancement in NLS1 hosts suggests that this gas simultaneously fuels both the central black hole and the surrounding stellar population. As the gas supply diminishes or feedback regulates inflow, the accretion 
rate declines and the source transitions toward a BLS1-like spectral state; orientation effects may further modulate where individual sources fall relative to the FWHM = 2000~km~s$^{-1}$ boundary. In this picture, the host galaxy sets the conditions for activity through its gas reservoir, while the instantaneous accretion state---potentially convolved with BLR inclination---determines the observed spectroscopic classification.

\section{Conclusion} \label{sec:conclusion}

Using CIGALE-based host--AGN decomposition for a sample of $\sim$12,000 broad-line AGNs from SDSS, we have investigated the physical origin of the NLS1/BLS1 dichotomy. Our main conclusions are as follows.

\begin{enumerate}

\item Across the two-dimensional FWHM(\Hb)--$L_{\rm AGN}$ plane, Fe~II emission strength ($R_{4570}$) varies primarily along lines of constant $\lambda_{\rm Edd}$, while global host-galaxy properties ($M_*$, SFR, $A_V$) follow lines of constant $M_{\rm BH}$ or $L_{\rm AGN}$. This behavior is more consistent with Eigenvector~1 being driven primarily by accretion state than by host mass alone.

\item In a mass-matched subsample of 767 NLS1--BLS1 pairs with statistically indistinguishable $M_*$, $M_{\rm BH}$, and $z$, NLS1s retain higher $\lambda_{\rm Edd}$, enhanced Fe~II emission, and bluer optical continua, showing that the NLS1 phenomenon is not explained by host stellar mass or virial black hole mass alone.

\item Residual differences in SFR, sSFR, and dust attenuation persist in the matched sample, indicating that host-galaxy conditions are not entirely negligible. This is consistent with a secondary role for host gas supply in modulating both star formation and nuclear fueling.

\item The FWHM(\Hb) $= 2000$~km~s$^{-1}$ boundary is not physically fundamental: host-galaxy properties and narrow-line ratios vary continuously across it, with no sharp discontinuity apparent in our diagnostics. The boundary nonetheless retains observational utility as a probabilistic selector of the high-$\lambda_{\rm Edd}$ regime.

\item NLS1s do not constitute a wholly distinct host-galaxy type or isolated AGN population, but rather represent the high-accretion end of a continuous Type 1 AGN sequence. The NLS1 spectral state likely reflects a high-accretion phase regulated by gas supply, with BLR inclination providing additional modulation of the observed line width.

\end{enumerate}

\begin{acknowledgments}
This work is supported by the China Manned Space Project (grant No.\ CMS-CSST-2025-A18) and the National Natural Science Foundation of China (NSFC; grant No.\ 12233005).
\end{acknowledgments}

\bibliography{references}{}
\bibliographystyle{aasjournalv7}

\appendix
\setcounter{figure}{0}            
\renewcommand{\thefigure}{A\arabic{figure}} 

\begin{figure}[!htbp]
    \centering
    \includegraphics{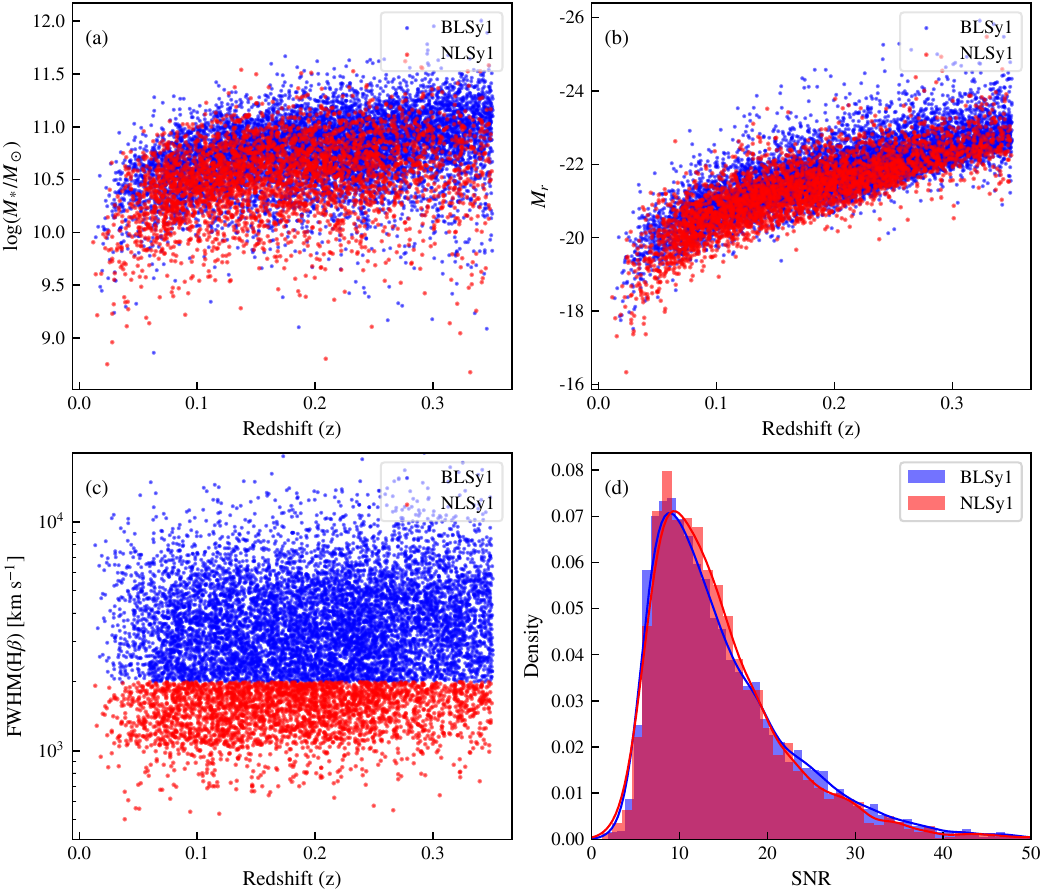}
    \caption{Bias assessment for the NLS1 (red) and BLS1 (blue) subsamples. Panels (a)-(c) show plots of redshift versus $\log M_*$, $M_r$, and $\mathrm{FWHM}(\mathrm{H}\beta)$. Panel (d) shows the spectral S/N distributions. Both populations cover similar redshift ranges ($z < 0.35$) with comparable distributions in $\log M_*$, $M_r$, $\mathrm{FWHM}(\mathrm{H}\beta)$, and spectral quality.}
    \label{fig:bias}
\end{figure}

\begin{figure}
    \centering
    \includegraphics[width=0.75\linewidth]{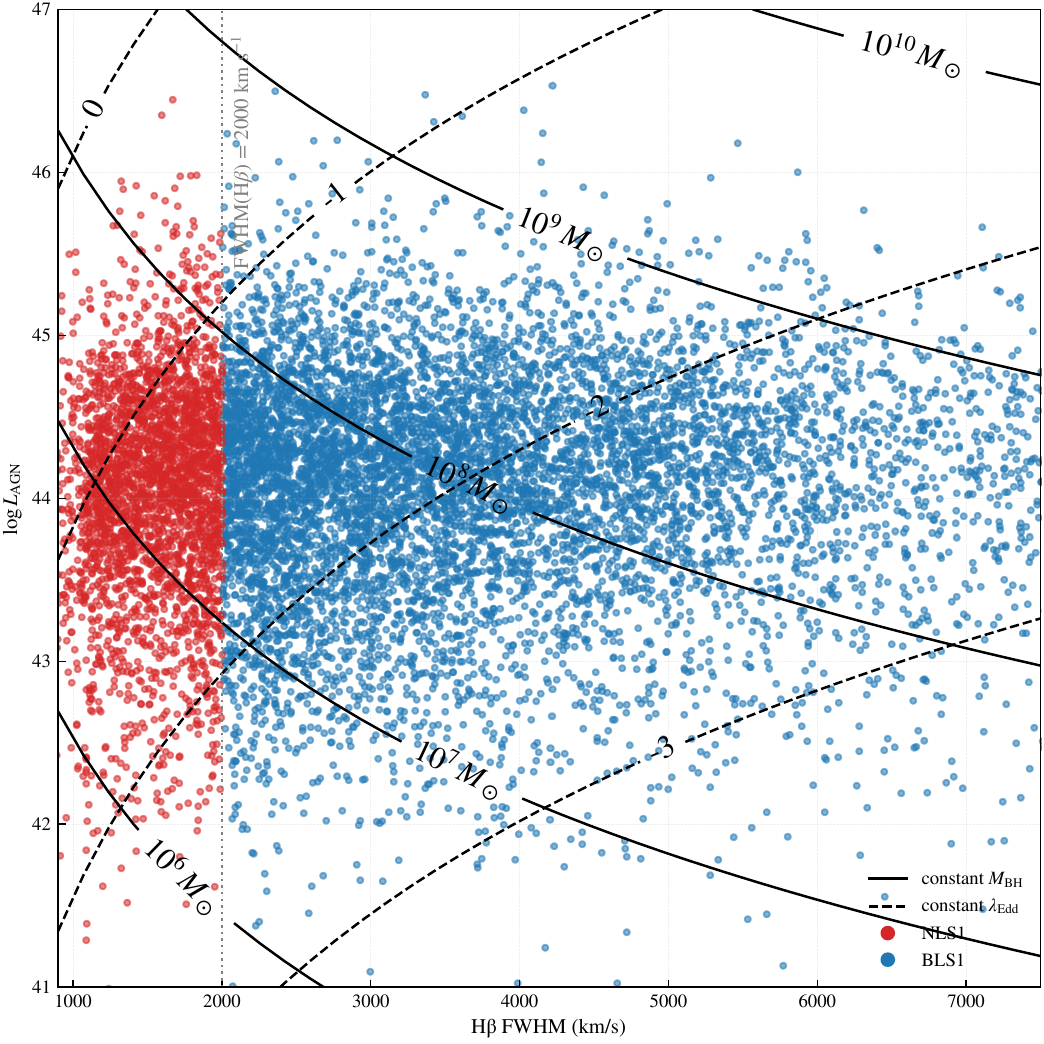}
    \caption{Distribution of the NLS1 and BLS1 samples in the FWHM(H$\beta$)-$L_{\rm AGN}$ plane. NLS1s and BLS1s are shown as red and blue points, respectively. Black solid contours indicate constant black hole mass, while black dashed contours indicate constant Eddington ratio. The vertical dotted line marks the conventional NLS1/BLS1 division at $\mathrm{FWHM}(\mathrm{H}\beta)=2000~\mathrm{km~s^{-1}}$. This figure illustrates that the two classes occupy overlapping regions of a continuous parameter space, with no obvious physical discontinuity at the empirical linewidth boundary. Instead, the distribution is more naturally organized by the underlying black hole mass and accretion state traced by the contour families.}
    % \label{fig:placeholder}
\end{figure}

\begin{figure}
    \centering
    \includegraphics[width=\linewidth]{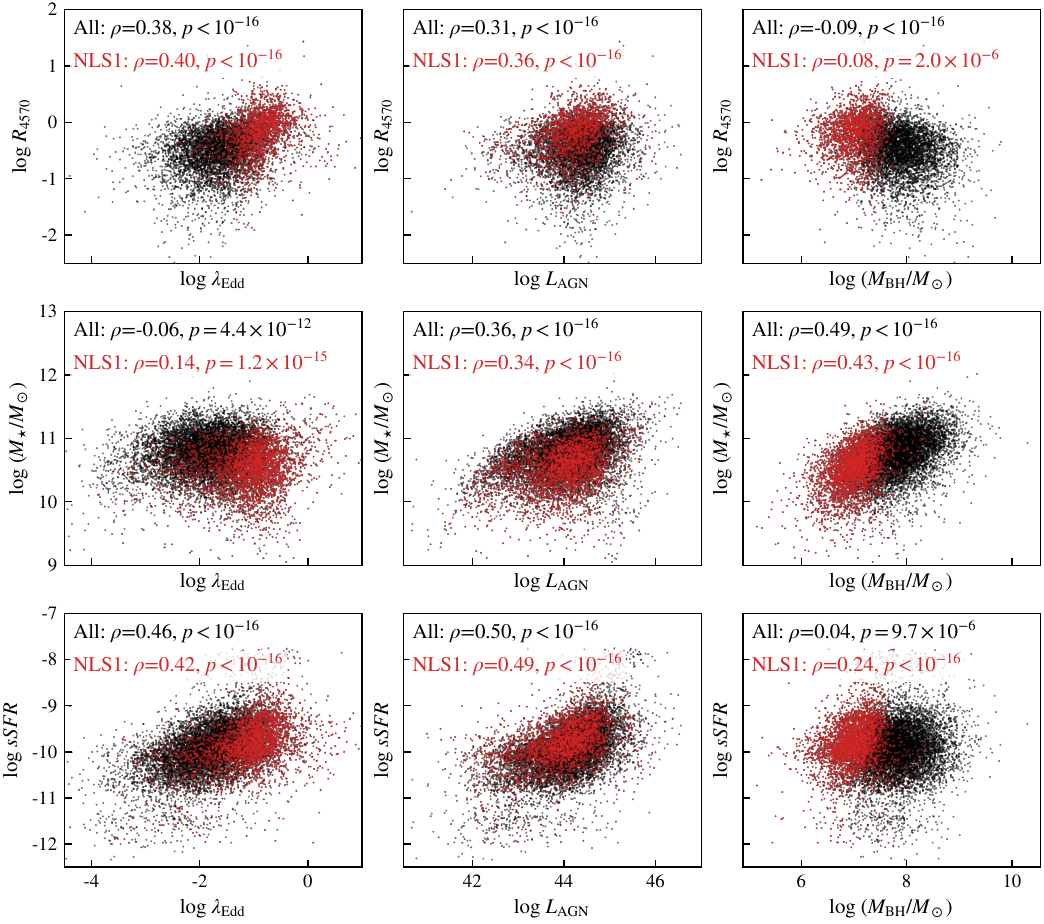}
    \caption{Comparison of the correlations of three key quantities with Eddington ratio, AGN luminosity, and black hole mass. From top to bottom, the panels show Fe~II strength ($R_{4570}$), host-galaxy stellar mass ($M_*$), and specific star formation rate (sSFR) as functions of $\lambda_{\rm Edd}$, $L_{\rm AGN}$, and $M_{\rm BH}$. Red and blue points denote NLS1s and BLS1s, respectively. The Spearman rank coefficients are reported for the full combined sample and, separately, for the NLS1 subsample. The full-sample coefficients highlight the behavior of the continuous Type 1 AGN sequence, while the NLS1 coefficients show whether the same trends remain within the narrow-line population alone. In this comparison, $R_{4570}$ shows a clearer preference for $\lambda_{\rm Edd}$ over $L_{\rm AGN}$ and $M_{\rm BH}$ than do the other quantities shown here.}
    \label{fig:corrs_comp}
\end{figure}

\begin{figure}[!htbp]
    \centering
    \includegraphics[width=0.9\textwidth]{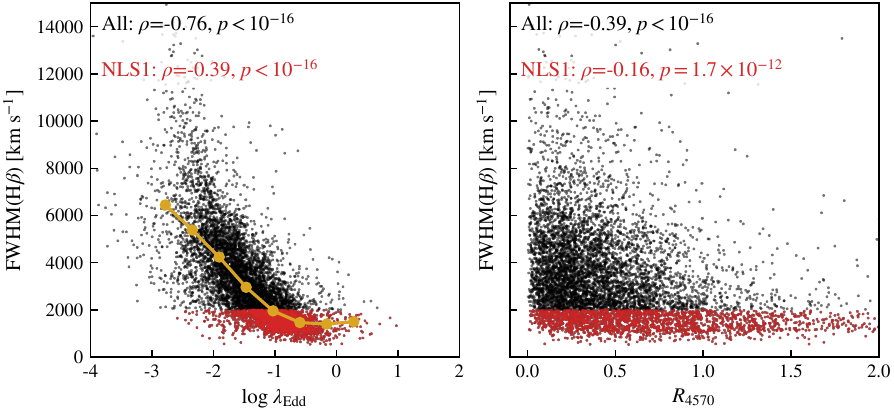} 
    \caption{The relationships between broad line kinematics, accretion state, and Eigenvector 1. Left: FWHM(\Hb) versus Eddington ratio ($\lambda_{\rm Edd}$). Right: FWHM(\Hb) versus Fe~II strength ($R_{4570}$). NLS1s are marked in red, while the underlying black points represent the full broad-line AGN sample. Spearman rank correlation coefficients ($\rho$) and corresponding $p$-values are annotated for both the combined sample and the NLS1 subsample alone. The significant degradation of the correlation in the right panel when restricted to NLS1s demonstrates that the 2000 $\mathrm{km\ s^{-1}}$ boundary artificially truncates a continuous physical sequence.}
    \label{fig:fwhm_correlations}
\end{figure}

\begin{figure}
    \centering
    \includegraphics[width=\linewidth]{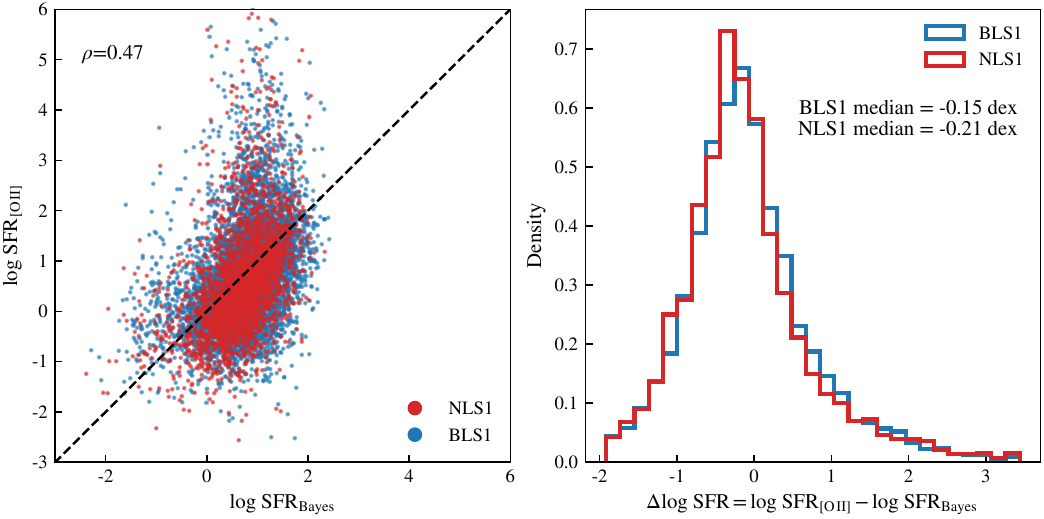}
    \caption{Comparison between star formation rates derived from extinction-corrected \OII luminosities and from CIGALE SED fitting. The \OII-based SFRs are computed following \citet{zhuang2019recalibration}, including a correction for the AGN narrow-line contribution estimated from \OIII and a metallicity correction based on the host stellar mass. Left: comparison of $\log {\rm SFR}_{[\mathrm{O\,II}]}$ and the CIGALE-based $\log {\rm SFR}_{\rm Bayes}$ for BLS1s (blue) and NLS1s (red); the dashed line marks the one-to-one relation. Right: distributions of $\Delta \log {\rm SFR} = \log {\rm SFR}_{[\mathrm{O\,II}]} - \log {\rm SFR}_{\rm Bayes}$ for the two populations. Although the two SFR estimates show some degree of scatter, as expected given their different spatial scales and underlying assumptions, the offset distributions of NLS1s and BLS1s are very similar. This suggests that the main comparative results are unlikely to be dominated by a differential bias in the adopted SFR estimator between the two populations.}
    % \label{fig:placeholder}
\end{figure}

\begin{figure}
    \centering
    \includegraphics[width=\linewidth]{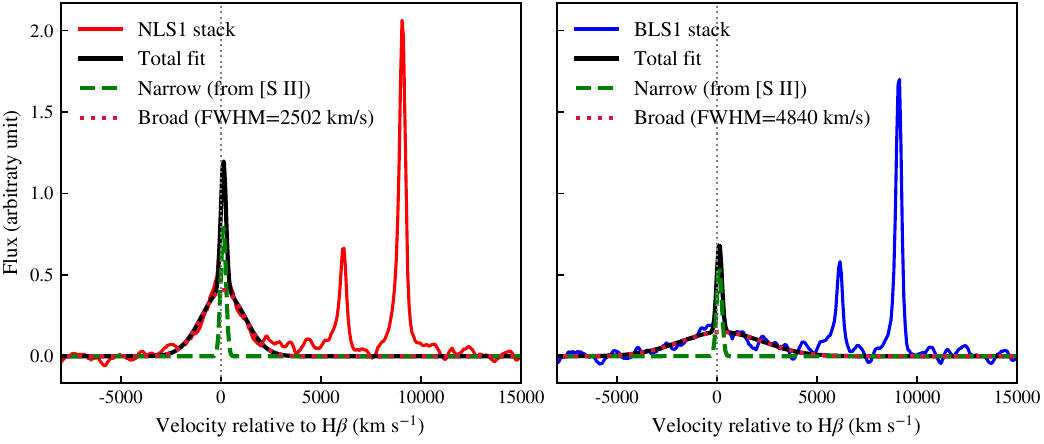}
    \caption{Decomposition of the stacked \Hb\ region in velocity space for the matched NLS1 and BLS1 composites. The continuum-subtracted spectra are shown together with the best-fitting total model and the individual narrow and broad \Hb\ components. The fitting is performed over an extended velocity range to capture the broad wings, and the narrow \Hb\ component is constrained using the [S~II] line profile.}
    \label{fig:hb_fit}
\end{figure}

\end{document}